\documentstyle[aps,twocolumn,psfig,epsfig,floats,citesort]{revtex}

\newcommand{\ageq}{\mbox{\
\raisebox{-.9ex}{$\stackrel{\textstyle >}{\sim}$}\ }}

\def\x{{\mbox{\boldmath$x$}}}

\def\eps{{\epsilon}}

\def\begineq{\begin{equation}}
\def\endeq{\end{equation}}
\def\be{\begin{equation}}
\def\ee{\end{equation}}

\begin{document}
\bibliographystyle{prsty}

\title{
The ultimate state of thermal convection
}

\author{Detlef Lohse$^1$ and Federico Toschi$^2$}
\address{
$^1$ Department of Applied Physics 
and J.\ M.\ Burgers Centre for
Fluid Dynamics,
University of Twente,\\
 P.O.Box 217, 7500 AE Enschede, Netherlands\\
$^2$ CNR, Istituto per le Applicazioni del Calcolo, Viale del Policlinico 137, I-00161 Roma, Italy and\\
INFM, Unit\`a di Tor Vergata, Via della Ricerca Scientifica 1, I-0
0133 Roma, Italy
}

\date{\today}

\maketitle

\begin{abstract}
The ultimate regime of thermal convection, the so called
Kraichnan regime (R.~H. Kraichnan, Phys. Fluids {\bf 5},  1374  (1962)),
 hitherto has been elusive. 
Here, numerical evidence for that regime is presented by performing
simulations  of the bulk of turbulence only, eliminating 
the thermal and kinetic boundary layers and replacing them by
periodic boundary conditions. 
\end{abstract}

\vspace{0.6cm}

Thermally driven turbulence is of tremendous importance 
in oceanography, geophysics, meteorology, astrophysics, or
process technology. 
If the flow is confined to a box
heated from below and cooled from above, the setup is called
Rayleigh-B\'enard convection. It is  
one of the classical problems 
in fluid dynamics \cite{kad01,cha81}. 
The control parameter are 
the dimensionless temperature difference between 
bottom and top (called Rayleigh number $Ra$), the ratio
between kinematic viscosity $\nu$ and heat diffusivity $\kappa$ (called
 Prandtl number $Pr$),
and the aspect ratio of the cell. The system answers
with some 
heat transfer 
from the bottom to the top (called Nusselt number $Nu$ in dimensionless form)
and the degree of turbulence in the cell (expressed in terms of the
Reynolds number $Re$).

All classical experiments on this system show (effective) power 
law dependences with $Nu \sim Ra^{0.25}$ up to $Nu \sim Ra^{0.33}$. 
However, for very
large $Ra$, Kraichnan \cite{kra62}
has predicted an ultimate scaling law
$Nu\sim Ra^{1/2}$. 
In this ultimate state the heat flux and the turbulent velocity
are expected  to be independent of the viscosity and heat diffusivity.
Indeed, for a system with infinite aspect ratio 
a mathematically strict upper bound $Nu \le 0.167 Ra^{1/2}-1$ could be 
found
\cite{doering}.

Hitherto, this ultimate state of thermal
convection remained elusive \cite{som99}, inspite of tremendous efforts
to find it. 
Niemela et al.\ \cite{nie00}
measured  up to $Ra\approx 10^{18}$
in helium gas close to the critical point ($Pr\approx 1$),
still finding an effective power law $Nu \sim Ra^{0.31}$. 
Glazier et al.\ \cite{gla99} measured up to $Ra\approx 10^{11}$ 
in mercury ($Pr=0.025$), finding $Nu \sim Ra^{0.28}$. 
The only evidence for a transition (at $Ra\approx 10^{11}$)
towards the ultimate regime 
has been claimed by Chavanne et al.\ \cite{cha97,cha01},
but Niemela and Sreenivasan \cite{nie02} 
argue that those data would be consistent
with $Nu\sim Ra^{1/3}$. 
In summary, it is therefore not clear 
\begin{itemize}
\item
whether in 
 today's experiments  the Rayleigh number is only still too small 
for the ultimate state, or
\item whether it does not exist at all. 
\end{itemize}

However, it is crucial to know 
which of these two alternative is the correct one: 
\begin{itemize}
\item
for practical reasons in view of above mentioned applications of thermal
convection in 
geophysics, industry etc, as upscaling from lab-scale models does not
work if there is a transition to a new state of turbulence;
\item
for fundamental reasons, as is a ``central dogma in  turbulence'' \cite{som99}
that its effects become independent of viscosity and diffusivity
  for large enough
Reynolds (or here Rayleigh) numbers \cite{fri95}. 
\end{itemize}
Indeed, also within Grossmann and Lohse's unifying theory on thermal
convection \cite{gro00,gro02} the Kraichnan or ultimate 
regime follows, once the thermal and kinetic boundary layers either
break down due to their expected instability at large $Ra$, or 
hardly contribute to the global kinetic and thermal dissipation
\cite{regimi}. 
According to ref.\ \cite{gro02} the breakdown of the laminar kinetic boundary
layer should happen at 
$Ra\approx 10^{14}$ for $Pr=1$ and at 
$Ra\approx 10^{12}$ for $Pr=0.025$ for an aspect ratio $\Gamma =1$ cell; 
Niemela and Sreenivasan 
\cite{nie02} find similar
values. Then 
the total 
 kinetic dissipation rate $\eps_u$ equals the bulk kinetic dissipation
$\eps_{u,bulk}$ and 
the total 
 thermal dissipation rate $\eps_\theta$ equals the bulk thermal dissipation
$\eps_{\theta,bulk}$ and the Kraichnan or ultimate regime with 
\be
Nu\sim Ra^{1/2} Pr^{1/2}
\label{k1}
\ee and 
\be 
Re\sim Ra^{1/2} Pr^{-1/2}
\label{k2}
\ee results \cite{gro00}.

From the above it follows that an artificial destruction of the boundary
layers should enhance the ultimate regime. Indeed, by performing
Rayleigh-B\'enard convection experiments in a cell with rough 
top and bottom walls Roche et al.\ \cite{roc01} find evidence for
the onset of the Kraichnan regime.

\begin{figure}[htb]
\setlength{\unitlength}{1.0cm}
\begin{picture}(11,11)
\put(0.0,11.5){\large (a)}
\put(0.0,5.6){\large (b)}
\put(0.5,11.5)
{\epsfig{figure=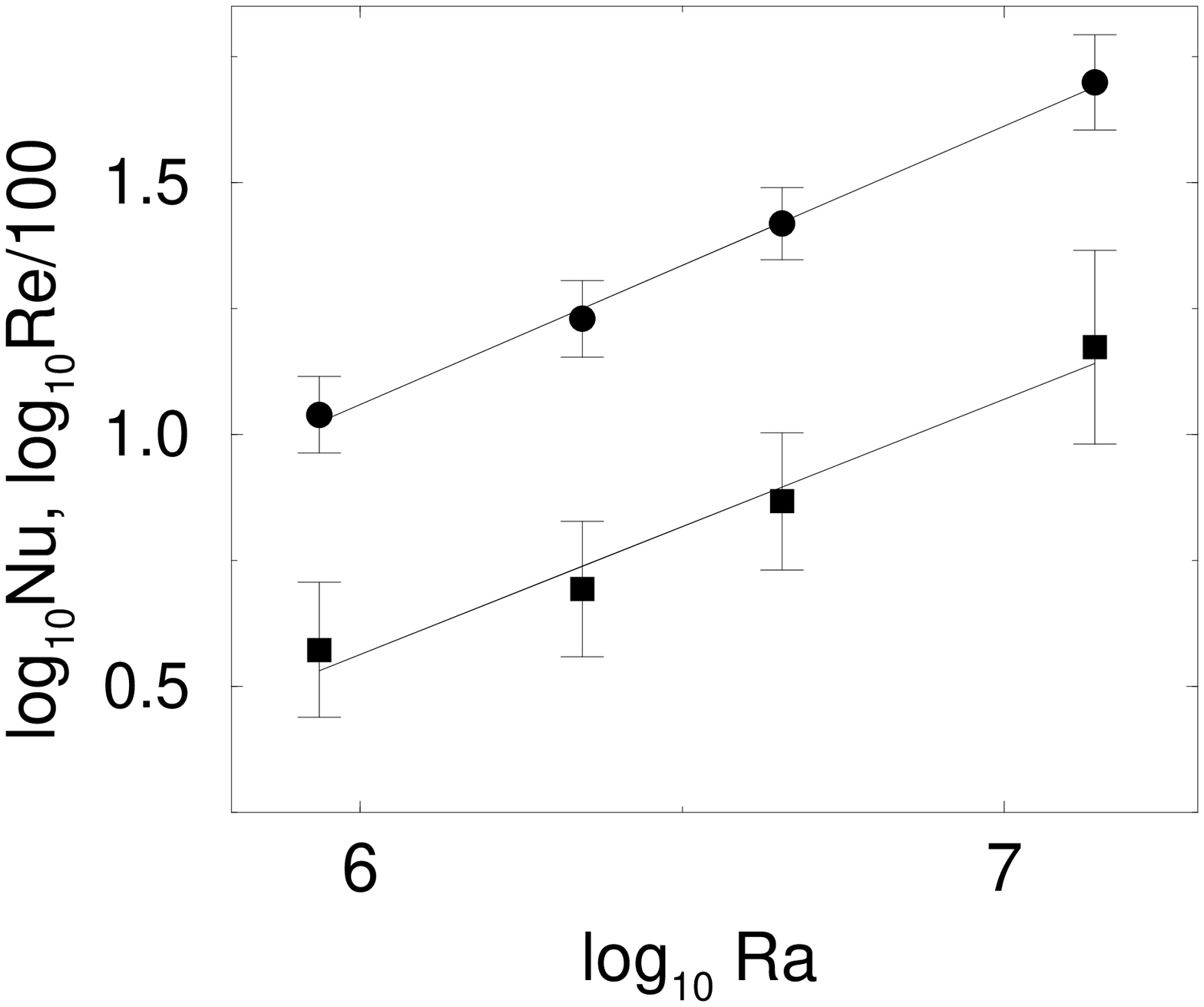,width=5.5cm,angle=-90}}
\put(0.5,5.6)
{\epsfig{figure=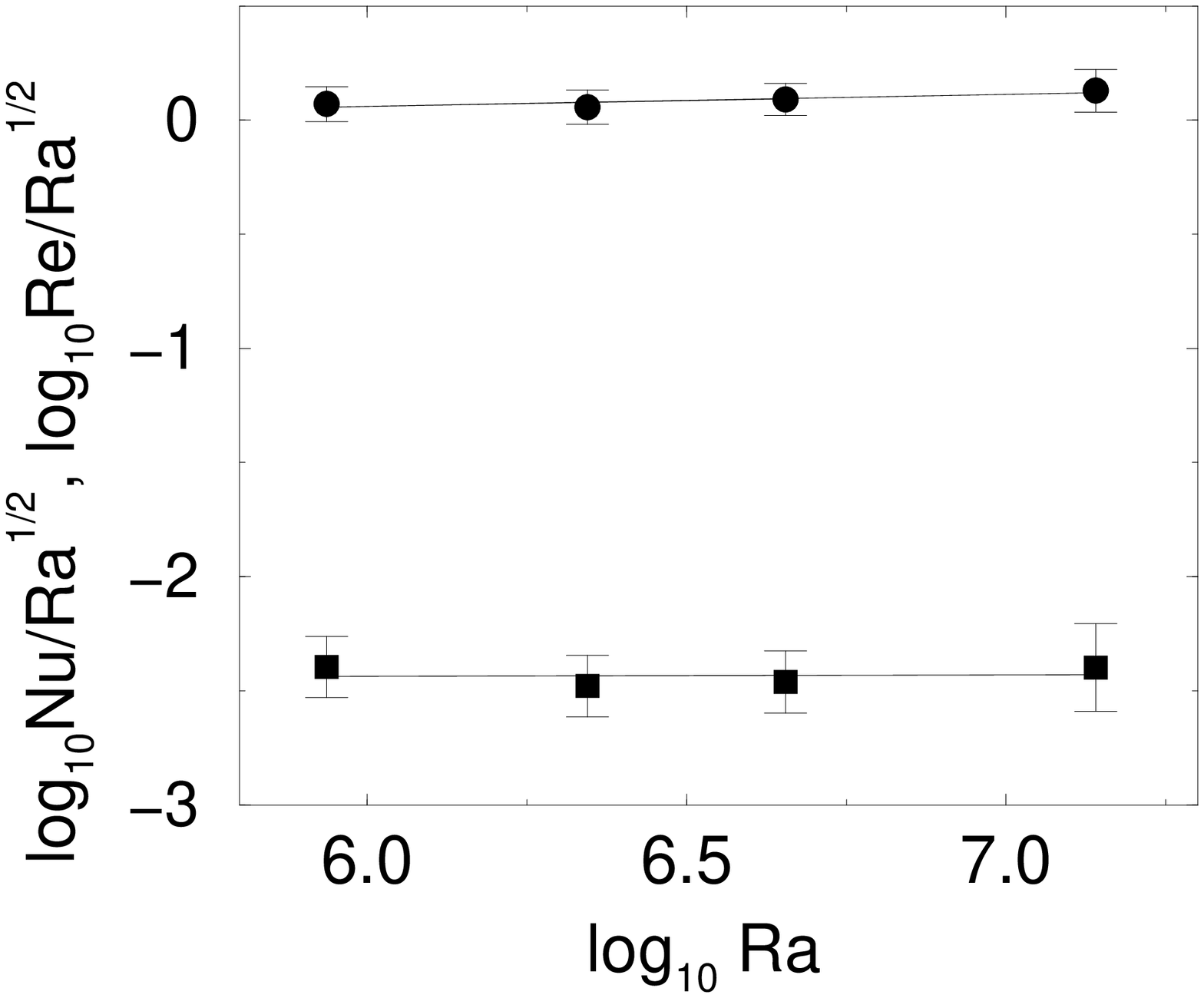,width=5.7cm,angle=-90}}
\end{picture}
\caption[]{
(a)
$Nu$ (squares, lower curve) and $Re$ (circles, upper curve) 
as a function of $Ra$ from our simulation.
The error bars are based on the variances of measured quantities
which have been obtained by splitting the total time series in
smaller pieces. Note that in this way we could also check that
our simulations are statistically stationary. 
The straight lines are linear regressions giving 
$Nu \sim Ra^{0.51\pm 0.06}$ and 
$Re \sim Ra^{0.55\pm 0.02}$.\\
(b) Same as in (a), but now compensated by $Ra^{1/2}$. 
}
\label{fig1}
\end{figure}

Here, we follow another approach: We perform numerical simulation {\it of the
bulk of turbulence only}, eliminating the thermal and kinetic boundary layers 
and replacing them by periodic boundary conditions \cite{tos}. 
Then the Kraichnan regime should follow immediately. 
The thermal bulk turbulence is forced by a mean temperature profile
$-z \Delta /L$, where
$L$ is the periodicity length and $\Delta $ the temperature difference
on that length scale. 
 The temperature fluctuations 
$  \theta$ then obey the advection equation
\be
\partial_t \theta + u_i \partial_i \theta = \kappa \partial_i \partial_i
 \theta + u_3 {\Delta \over L}. 
\label{ad}
\ee
The velocity field $u_i(\x , t)$ obeys the standard Boussinesq equation,
\be
\partial_t u_i + u_j \partial_j u_i = -\partial_i p +\nu \partial_j^2 u_i
+\beta g \delta_{i3} \theta .
\label{bou}
\ee
$\beta$ is the thermal expansion coefficient, $g$ gravity, 
and $p$ the
 pressure. 
$Ra$ and $Pr$ are
 defined as usual as $Ra = \beta g \Delta L^3 /(\nu \kappa )$
and $Pr=\nu/\kappa$, respectively. The Prandtl number is 1 in our
simulations presented here. 
Also $Nu$ is defined as usual as 
$Nu = \left< u_3 \theta \right> L / (\kappa \Delta ) - \partial_3
\left< \theta \right>_A L/\Delta$.
The Reynolds number is defined with the rms mean velocity fluctuation 
$u'$, i.e., $Re = u' L/\nu$. 
A similar simulation has earlier been performed by 
Borue and Orszag \cite{borue}. The focus of that work was on scaling
properties and spectra and thus hyperviscosity has been employed.

From eqs.\ (\ref{ad}) and (\ref{bou}) 
one can derive by volume averaging and 
assuming statistically stationarity the exact \cite{exact}
relations 
$\eps_\theta = \kappa L^{-2}  \Delta^2 Nu$ and
$\eps_u = \nu^3 L^{-4} Nu Ra Pr^{-2}$. 
As in refs.\ \cite{gro00} we estimate 
$\eps_{u,bulk} \sim u^{\prime 3}/L$ and 
$\eps_{\theta,bulk} \sim u'\Delta^2/L$ and the Kraichnan regime
(\ref{k1}) -- (\ref{k2}) results. 
Note that $L$ is the relevant length scale 
for these estimates as for the Bolgiano length
scale it holds $l_{B} \ageq L$.

This Kraichnan scaling is indeed seen in our numerical simulations
of eqs.\ (\ref{ad}) and (\ref{bou}).
We run four different $Ra$ numbers 
$Ra= 8.64 \cdot 10^5$ (for 160 large eddy turnovers),
$Ra= 2.21 \cdot 10^6$ (for 60), 
$Ra= 4.51 \cdot 10^6$ (for 180), and 
$Ra= 1.38 \cdot 10^7$ (for 334), and average over space and time to
obtain $Nu$ and $Re$ \cite{ss}. 
The $Ra$-scaling is consistent with 
eqs.\ (\ref{k1}) -- (\ref{k2}),
see figure \ref{fig1}. 
To our knowledge it is the first realization of the ultimate 
regime of thermal convection. It is remarkable that it is realized
in spite of the relatively low $Ra$ numbers of the simulations. 
The reason is that the boundary layers have been eliminated and
the simulations focus on the bulk.
Due to the lack of boundary layers
and the more efficient driving,
the Reynolds numbers
achieved here are much larger than they would be in the standard
Rayleigh-B\'enard case. Vice versa, the required Rayleigh numbers
to achieve the same degree of turbulence for the standard Rayleigh-B\'enard
case as here are much higher. E.g., to achieve $Re=4996$ (here for
$Ra= 1.38 \cdot 10^7$) one needs a Rayleigh number of 
$Ra = 7.2 \cdot 10^8$ in the standard case \cite{nie01}.

In conclusion, from above alternative the first one is favorable:
Once $Ra$ is so large that the laminar kinetic boundary layers breaks down, 
the ultimate regime of thermal convection,
which has been so elusive in experiment, should finally show up.
Obviously, the ultimate proof for the existence of this regime can only
come from experiment. 

\noindent
{\it Acknowledgement:}
The work is part of the research  program of the Stichting voor 
Fundamenteel Onderzoek der Materie (FOM), which is financially supported 
by the Nederlandse  Organisatie voor Wetenschappelijk Onderzoek (NWO).
We acknowledge support from
the European Union (EU) through the
European Research Network on ``Nonideal Turbulence''
(contract HPRN--CT--200000162). We also acknowledge INFN, 
Sezione di Pisa for the APEmille computer resources.\\


\begin{thebibliography}{10}
\bibitem{exact}
Note that
for  the standard RB case the corresponding 
second relation 
reads $\eps_u = \nu^3 L^{-4} (Nu-1) Ra Pr^{-2}$.
\end{thebibliography}

\end{document}